\documentclass{PoS}
\usepackage{amsmath}
\usepackage{siunitx}
\usepackage{url}
\usepackage{xcolor}
\usepackage{xspace}

\newcommand*\oline[1]{
  \hspace*{0.2em}
  \vbox{
    \kern-0.35ex
    \hrule height 0.4pt
    \kern0.35ex
    \hbox{
      \kern-0.5em
      \ifmmode#1\else\ensuremath{#1}\fi
      \kern-0.0em
}}}

\newcommand{\gev}{\text{\si{\giga\electronvolt}}\xspace}
\newcommand{\tev}{\text{\si{\tera\electronvolt}}\xspace}

\definecolorset{rgb}{}{}{fuchsia,0.54,0.39,1;dandelion,0.902,0.58,0.3;delphinium,0.60,0.59,1;rosered,0.75,0.35,0.35;royalblue,0.0,0.5,1.0;forestgreen,0.0,0.7,0.0}

\newcommand*\red[1]{\begin{color}{red}#1\end{color}}
\newcommand*\blue[1]{\begin{color}{royalblue}#1\end{color}}
\newcommand*\green[1]{\begin{color}{forestgreen}#1\end{color}}

\title{Fitting the two-loop renormalized Two-Higgs-Doublet model}

\ShortTitle{Fitting the two-loop renormalized Two-Higgs-Doublet model}

\author{\speaker{Otto Eberhardt}%
         \thanks{based on arXiv:1503.08216 \cite{Chowdhury:2015yja} in collaboration with Debtosh Chowdhury}
         \\
        Istituto Nazionale di Fisica Nucleare, Sezione di Roma\\
        E-mail: \email{otto.eberhardt@roma1.infn.it}}

\abstract{We present global fits to the Two-Higgs-Doublet model, assuming a
  softly broken $Z_2$ symmetry of the types I and II
  and CP conservation in the scalar potential.
  We show how much the parameter space is constrained
  by the combination of the relevant theoretical and experimental
  inputs,
  including the LHC data after the first run
  and interpreting the 125 GeV boson as the light CP even Higgs.
  Using the next-to-leading order renormalization
  group equations, we address the questions of vacuum stability
  and the hierarchy problem in the context of the mentioned Two-Higgs-Doublet
  models.}

\FullConference{18th International Conference From the Planck Scale to the Electroweak Scale \\
		 25-29 May 2015\\
		 Ioannina, Greece }

\begin{document}

\section{Introduction}

The theory community of particle physics agrees that their Standard Model (SM) has some shortcomings which are to be resolved in a
more complete theory, like for example supersymmetry (SUSY).
The Two-Higgs-Doublet model (2HDM) belongs to the most popular extensions of the SM and contains the Higgs sector of supersymmetry.
One of the problems of the SM is the instability of the Higgs self-coupling under renormalization group running \cite{Degrassi:2012ry,Buttazzo:2013uya}, which can be solved in a 2HDM or SUSY. Another SM issue for which SUSY offers an elegant solution is the naturalness of the Higgs mass $m_h=125$ \gev. In the 2HDM, however, there is no mechanism which accounts for the natural occurrence of a Higgs at electroweak scales and thus avoids fine-tuning of its mass. Earlier work has shown that a cancellation of quadratic divergencies, which is essential for a naturally light Higgs mass, is possible at leading \cite{Ma:2001sj} and next-to-leading order \cite{Grzadkowski:2009iz}; but no cancellation to all orders could be guaranteed.
Here, we want to derive the next-to-leading order renormalization group equations (NLO RGE) and analyse their effects on the stability of the Higgs potential as compared to the leading order (LO) results. We then want to make use of them to investigate whether a region in the 2HDM parameter space can be found for which a light $h$ mass occurs naturally and higher order terms quadratic in the cut-off scale are suppressed. In order to understand the complete picture we will rely on global fits rather than a couple of benchmark scenarios. The fits -- performed with the latest experimental constraints -- also revealed some novel features and quantify the maximal deviation from the alignment limit, in which the $125$ \gev Higgs looks like the SM one.

\section{The 2HDM with a softly broken $Z_2$ symmetry}

One obtains the 2HDM by the addition of a second Higgs doublet to the SM field content.
In order to avoid the appearance of flavour-changing neutral currents at tree-level, we add a $Z_2$ symmetry to the 2HDM, which can be broken softly. Hence, the Higgs potential for the two doublets $\Phi_1$ and $\Phi_2$ reads

\begin{align}
 V
 &=m_{11}^2\Phi_1^\dagger\Phi_1^{\phantom{\dagger}}
   +m_{22}^2\Phi_2^\dagger\Phi_2^{\phantom{\dagger}}
   -m_{12}^2 ( \Phi_1^\dagger\Phi_2^{\phantom{\dagger}}
              +\Phi_2^\dagger\Phi_1^{\phantom{\dagger}})
   +\tfrac12 \lambda_1(\Phi_1^\dagger\Phi_1^{\phantom{\dagger}})^2
   +\tfrac12 \lambda_2(\Phi_2^\dagger\Phi_2^{\phantom{\dagger}})^2
 \nonumber \\
 &\phantom{{}={}}
  +\lambda_3(\Phi_1^\dagger\Phi_1^{\phantom{\dagger}})
            (\Phi_2^\dagger\Phi_2^{\phantom{\dagger}})
  +\lambda_4(\Phi_1^\dagger\Phi_2^{\phantom{\dagger}})
            (\Phi_2^\dagger\Phi_1^{\phantom{\dagger}})
  +\tfrac12 \lambda_5 \left[ (\Phi_1^\dagger\Phi_2^{\phantom{\dagger}})^2
                      +(\Phi_2^\dagger\Phi_1^{\phantom{\dagger}})^2 \right],\label{eq:pot}
\end{align}

where the eight potential parameters are assumed to be real, excluding $CP$ violation in the Higgs sector. Switching to a physical basis, we can transform them into the following parameters: the masses of the physical Higgs states, that is $m_h$ and $m_H$ for the light and heavy $CP$-even scalars -- of which the former is interpreted as the $125$ \gev resonance seen at the LHC --, $m_A$ for the $CP$-odd scalar and $m_{H^+}$ for the charged scalars, two mixing angles $\alpha$ and $\beta$, the electroweak vacuum expectation value $v\approx 246$ \gev and $m_{12}^2$ from \eqref{eq:pot} as soft $Z_2$ breaking quantity. Since two of the physical parameters have been precisely measured, we will fix them to their central values and end up with six free parameters. For convenience we will rather use the angle combinations $\beta-\alpha$ and $\tan \beta$. The reason is the following: the precise measurement of the $h$ properties at LHC push the 2HDM parameters into a region in which the $h$ couplings to gauge bosons are SM-like. This is the case if $\beta-\alpha=\pi/2$, which is called the alignment limit of the 2HDM. It is different from the decoupling limit; the latter is a special case of the alignment limit for very heavy $m_H$, $m_A$ and $m_{H^+}$.

The mentioned $Z_2$ symmetry results in the fact that up-type and down-type quarks and charged leptons couple to only one of the two Higgs fields in \eqref{eq:pot} each. Without loss of generality we can attribute $\Phi_2$ to the up-type quarks, so four different possibilities remain.
Here, we will only present type I and II, where both, down-type quarks and charged leptons couple either only to $\Phi_2$ (type I) or only to $\Phi_1$ (type II); the results for the other two types can be found in \cite{Chowdhury:2015yja}. In the following, we will only consider the Yukawa couplings to $t$ and $b$ quarks and the $\tau$ lepton; the other contributions are negligible in the discussion of the RGE.

\section{NLO RGE}

\begin{figure}
  \centering
  \resizebox{200pt}{!}{
   \begin{picture}(300,300)(0,0)
    \put(-40,87){\includegraphics[width=340pt]{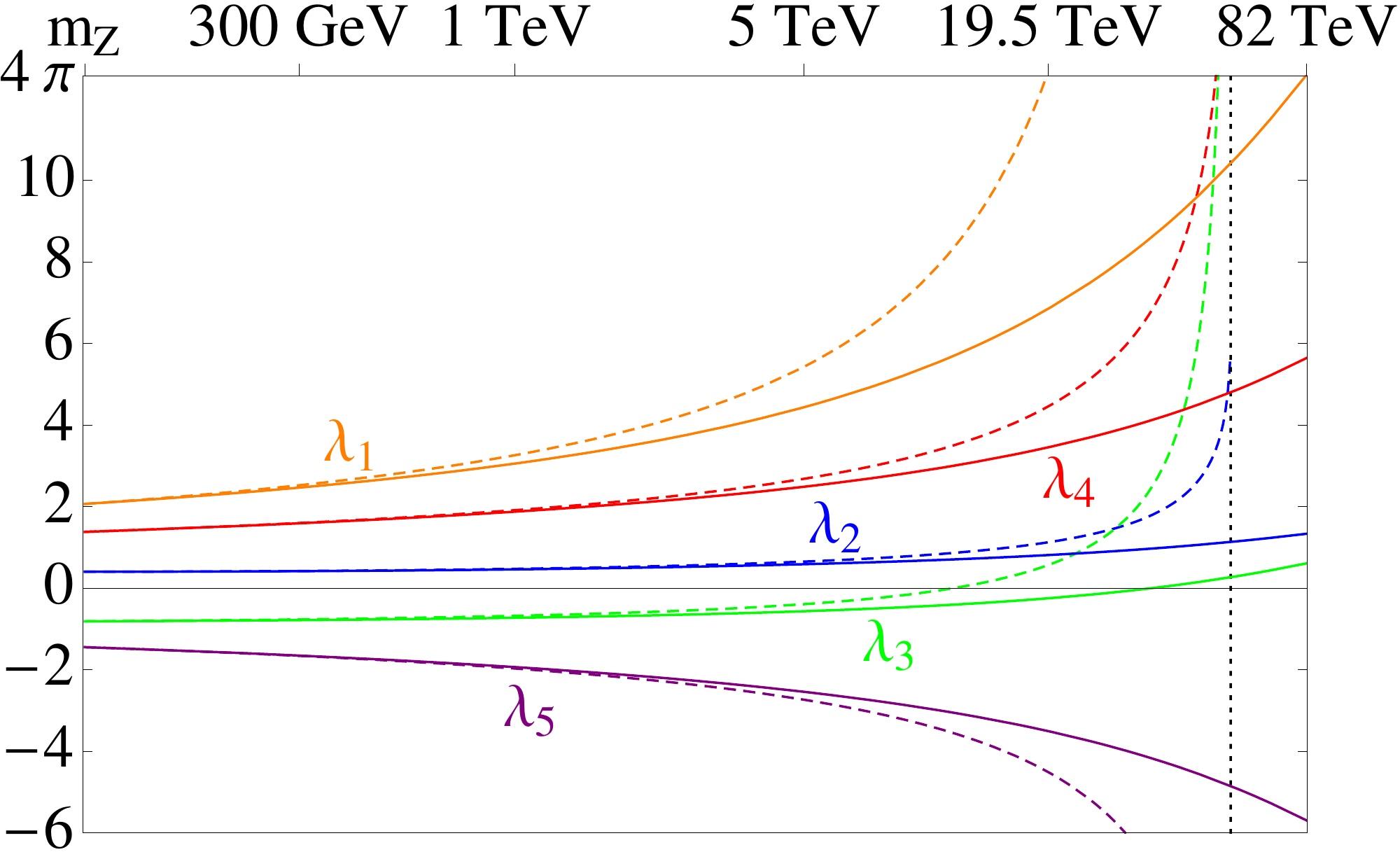}}
    \put(-55,-10){\includegraphics[width=354.5pt]{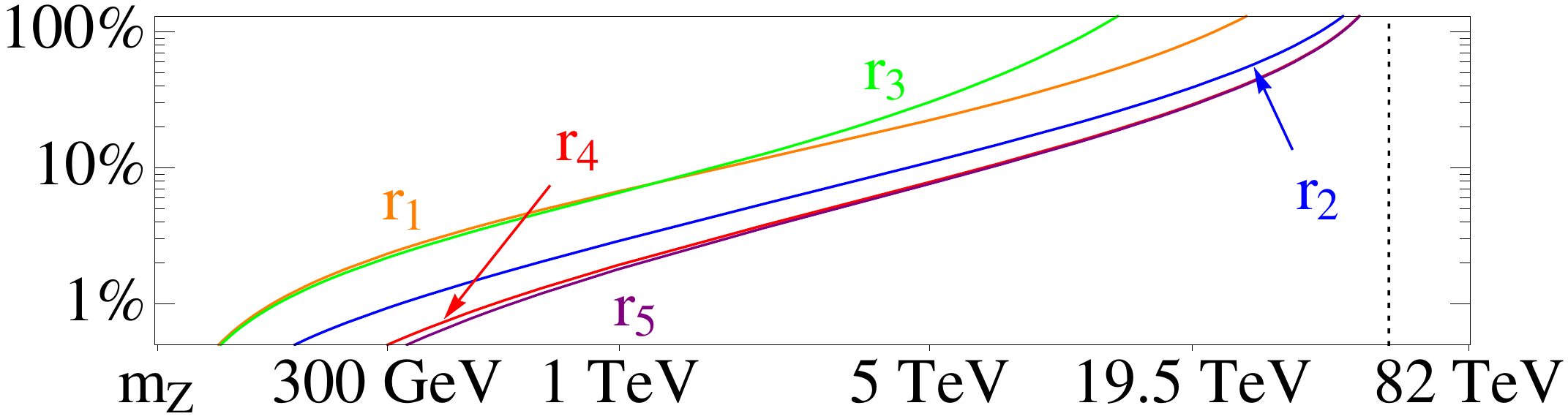}}
   \end{picture}
  }
  \resizebox{200pt}{!}{
   \begin{picture}(300,300)(0,0)
    \put(10,139){\includegraphics[width=346.5pt]{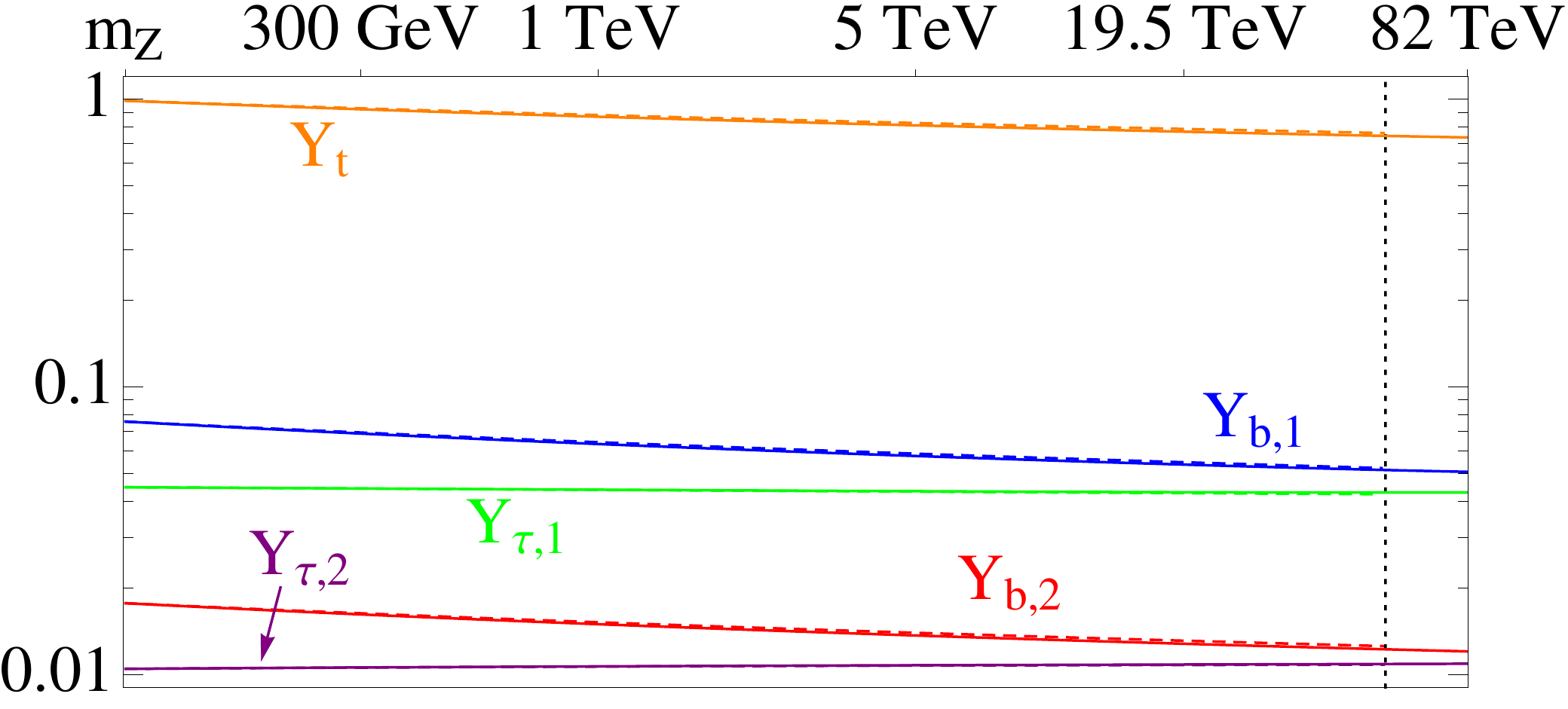}}
    \put(-16.5,-10.5){\includegraphics[width=373pt]{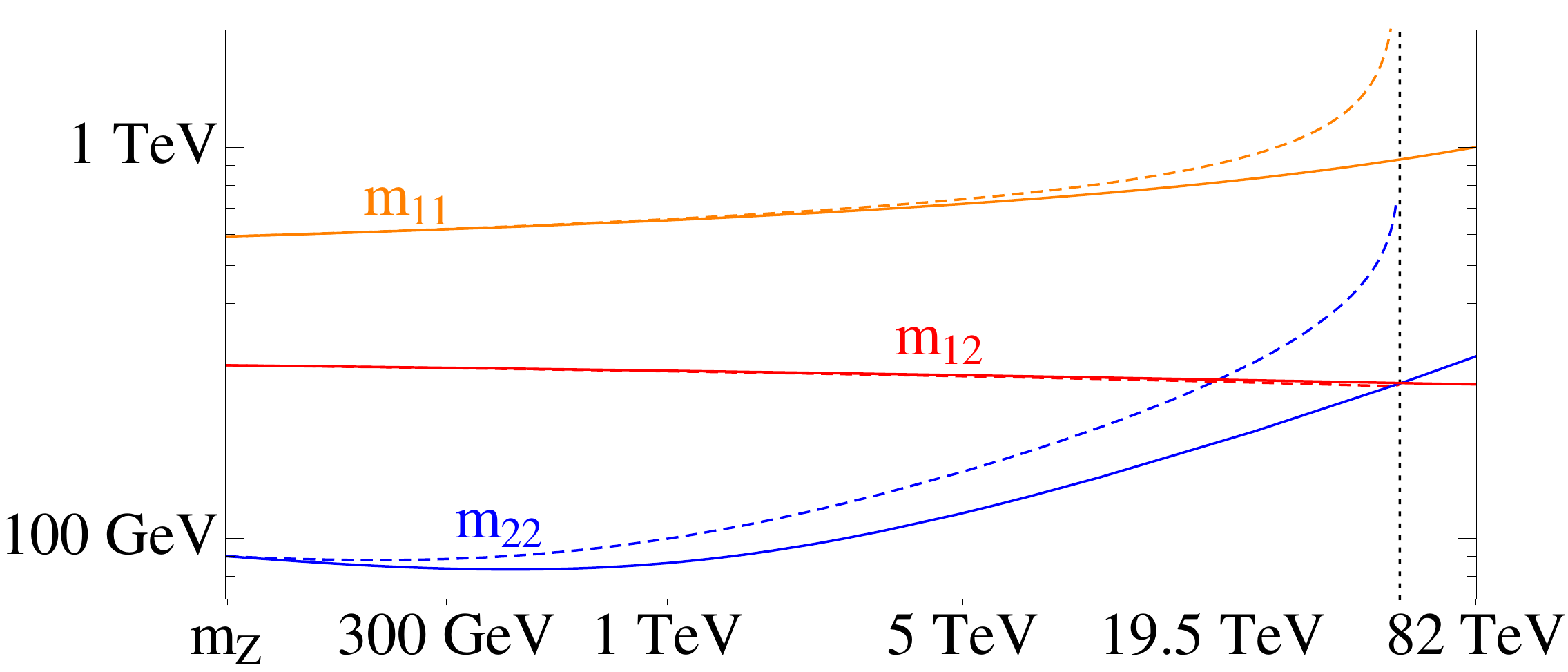}}
   \end{picture}
  }
  \caption{We show the LO (dashed) and NLO (solid) RG evolution for a bechmark scenario. The top left panel shows the running of the quartic couplings $\lambda_i$; the relative difference between LO and NLO expressions $r_i=\left| (\lambda _i^{\text{\tiny{LO}}}-\lambda _i^{\text{\tiny{NLO}}})/\lambda _i^{\text{\tiny{NLO}}} \right|$ can be seen in the bottom left panel. On the right, the running of the Yukawa couplings (top, $Y_{b,1}=Y_{\tau,1}\equiv0$ in type I and $Y_{b,2}=Y_{\tau,2}\equiv0$ in type II) and the quadratic potential parameters (bottom) is illustrated. While the LO RGE have a perturbative cut-off at $19.5$ \tev and a Landau pole at $54$ \tev (dotted line), the NLO corrections "stabilize" the running such that perturbativity only breaks down at $82$ \tev.}
  \label{fig:LOvsNLO}
\end{figure}

\begin{figure}
  \centering
  \resizebox{400pt}{!}{
   \begin{picture}(400,130)(0,0)
    \put(-40,0){\includegraphics[width=220pt]{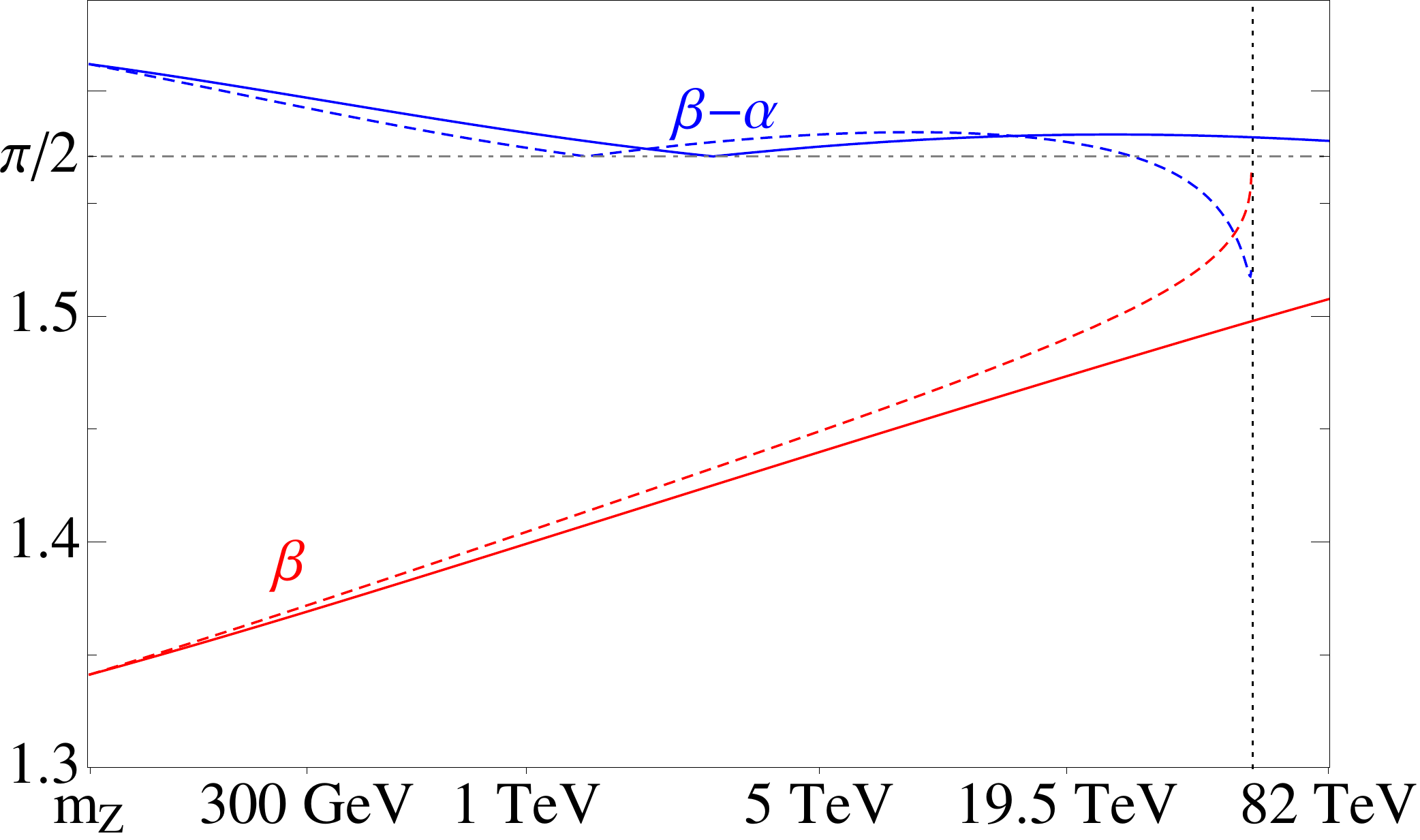}}
    \put(190,0.1){\includegraphics[width=241pt]{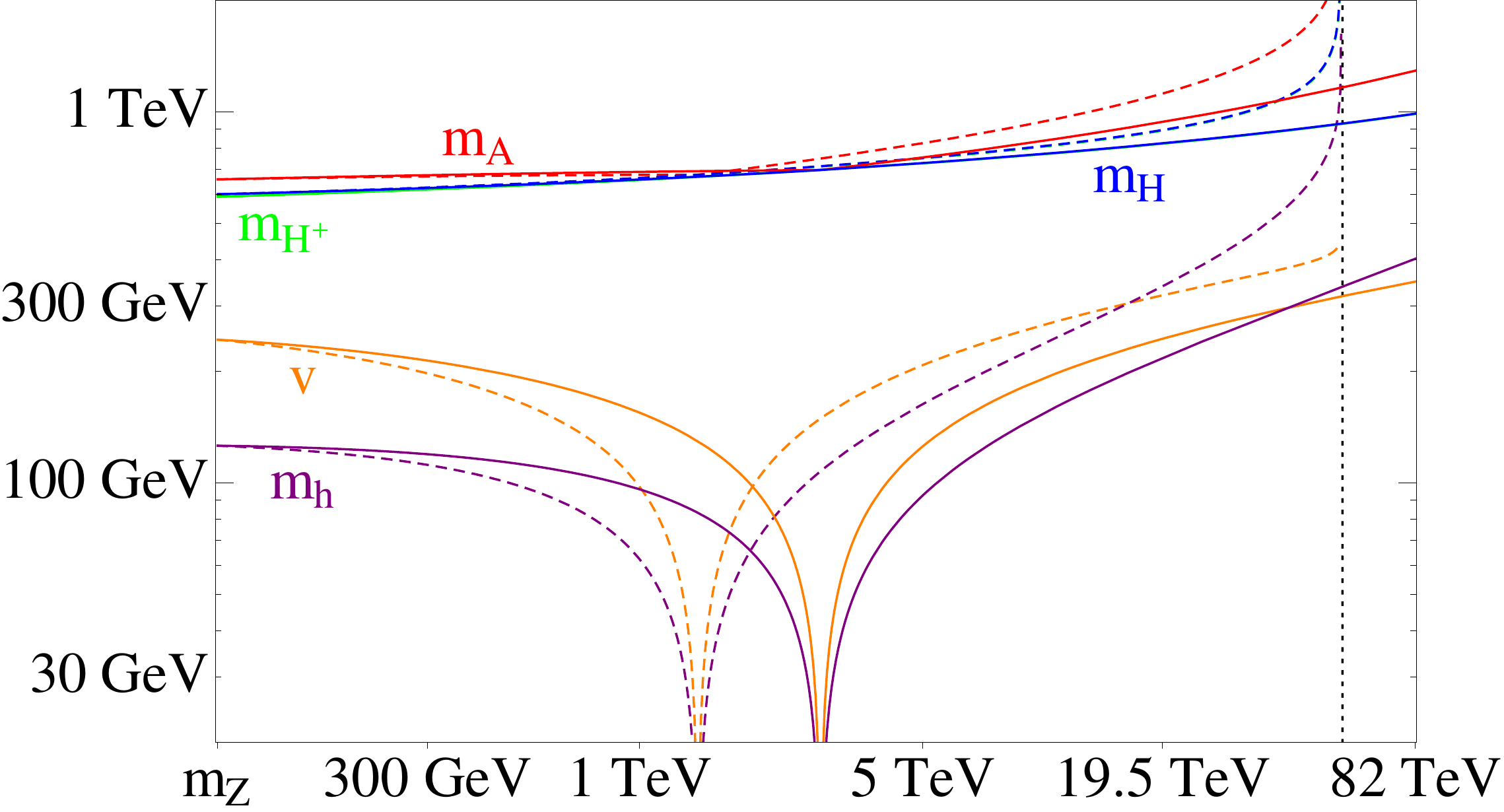}}
   \end{picture}
  }
  \caption{LO (dashed) and NLO (solid) RG evolution of the physical mixing angles and masses for the same benchmark scenario as in Fig.~\protect\ref{fig:LOvsNLO}. $H^+$ and $H$ are almost mass degenerate. We observe a breakdown of $v$ and $m_h$ for $\beta-\alpha=\pi/2$ at $1.4$ \tev ($2.8$ \tev) at LO (NLO).}
  \label{fig:LOvsNLOphys}
\end{figure}

We use the python code PyR@TE \cite{Lyonnet:2013dna} to obtain the renormalization group equations at NLO. The expressions for all four types of $Z_2$ can be found in \cite{Chowdhury:2015yja}. In order to compare to the leading order RGE, we pick the benchmark point H-4 from \cite{Baglio:2014nea} and show the evolution of the potential parameters (Fig.~\ref{fig:LOvsNLO}) and physical parameters (Fig.~\ref{fig:LOvsNLOphys}) up to the point at which the quartic couplings become non-perturbative (i.e. $\lambda_i>4\pi$). 

We chose this benchmark scenario (defined by $m_H=600$ \gev, $m_A=658$ \gev, $m_{H^+}=591$ \gev, $\beta-\alpha=0.513 \pi$, $\tan \beta=4.28$, and $m_{12}^2=76900\;\gev^2$), because in spite of relatively large quartic couplings at the electroweak scale $m_Z$, the perturbative breakdown only occurs at a scale far beyond LHC reach. The observation that the NLO corrections to the RGE shift this breakdown to higher scales as compared to the LO RGE was found to be generally valid for the 2HDM. Here the NLO perturbativity limit is even reached beyond the LO Landau pole (upper left panel of Fig.~\ref{fig:LOvsNLO}). A naive comparison between NLO and LO RGE for the quartic couplings is the relative ratio of their difference $r_i$, shown in the panel below, which can be of order 10\% for perturbative $|\lambda_i|<2\pi$. The $r_i$ have to be taken with a grain of salt, though, because they become large by definition if the NLO $\lambda_i$ becomes very small. In \cite{Chowdhury:2015yja} we also use the more stable relative distance between LO and NLO curves. In both cases, the NLO corrections can amount to order of 10\% effects in the perturbative regime.
The right hand panels of Fig.~\ref{fig:LOvsNLO} show the evolution of the Yukawa couplings and the massive potential parameters. While for all potential parameters the choice of the $Z_2$ symmetry has no visible impact on the RGE behaviour, it is important for the starting value of the Yukawa couplings to the $b$ quark and the $\tau$ lepton. In type I (II), $Y_{b,1}$ and $Y_{\tau,1}$ ($Y_{b,2}$ and $Y_{\tau,2}$) are zero by definition.
The running of the mass parameters in Fig.~\ref{fig:LOvsNLO} shows that also their magnitude can change significantly at higher scales.
If we switch to the physical basis, we observe in the left panel of Fig.~\ref{fig:LOvsNLOphys} that $\beta-\alpha$ runs into the alignment limit, which corresponds to a breakdown of $v$ and $m_h$ visible in the right panel.

After scrutinizing this benchmark scenario, we want to continue with the presentation of the fitting set-up, which is required to extend our analysis to the whole parameter space of the 2HDM types I and II.

\section{The fitting set-up}

In order to respect all relevant existing constraints on the 2HDM, we combine various theoretical and experimental inputs in a frequentist global fit, performed with the CKMfitter package \cite{Hocker:2001xe}:
Assuming that the $125$ \gev Higgs is the light Higgs $h$, our spectrum for the heavy Higgs particles $H$, $A$ and $H^+$ ranges from $130$ \gev to 10 \tev.
The Higgs potential should be bounded from below at all scales \cite{Deshpande:1977rw} and the electroweak vacuum $v$ is assumed to be the global minimum at $m_Z$ \cite{Barroso:2013awa}. Since the above assumption of perturbativity of the quartic Higgs couplings is parametrisation dependent, we rather demand the $S$-matrix of $\Phi_i\Phi_j\to\Phi_i\Phi_j$ processes to be unitary, which affects physical states and thus does not depend on the chosen parametrisation \cite{Ginzburg:2005dt}. As suggested by previous work (see the discussion in \cite{Baglio:2014nea}), we require that the $S$-matrix eigenvalues are smaller than $1/8$ instead of $1$ at all scales excluding extreme 2HDM scenarios with a breakdown at relatively low energy scales (see \cite{Chowdhury:2015yja} for details). When we run a parameter set to higher scales we will interrupt the RGE evolution either if one of the stability criteria (positivity or unitarity) is violated or if the Planck scale $\mu_{\text{Pl}}=10^{19}$ \gev is reached; we will call the cut-off scale $\mu_{\text{stability}}$.

For the type II, a new determination of the branching ratio of $\oline{B}\to X_s\gamma$ decays yields a lower bound on $m_{H^+}$ of $480$ \gev at the 95\% confidence level \cite{Misiak:2015xwa}. We also include the mass difference in the $B_s$ system $\Delta m_s$ \cite{Deschamps:2009rh}.
Since the oblique parameters $S$, $T$ and $U$ do not account for vertex corrections, we use the full set of electroweak precision observables (EWPO) measured at the LEP and SLD colliders, adding the 2HDM one-loop contributions \cite{Hahn:2000kx,Hahn:1998yk,Hahn:2006qw} to the Zfitter code \cite{Bardin:1989tq, Bardin:1999yd, Arbuzov:2005ma}.
$h$ signal strengths and direct heavy $H$ and $A$ searches are included with the help of HDECAY branching ratios \cite{Djouadi:1997yw,Butterworth:2010ym,Djouadi:2006bz}.
We also use recent $H^+$ searches, which however only affect extreme $\tan \beta$ regions. For the experimental references, we refer to our publication \cite{Chowdhury:2015yja}.
If not mentioned otherwise, exclusion statements are meant to be understood at the $2\sigma$ level, corresponding roughly to the 95\% confidence level.

\section{Results}

\begin{figure}
  \centering
  \resizebox{450pt}{!}{
   \begin{picture}(450,330)(0,0)
    \put(0,165){\includegraphics[width=220pt]{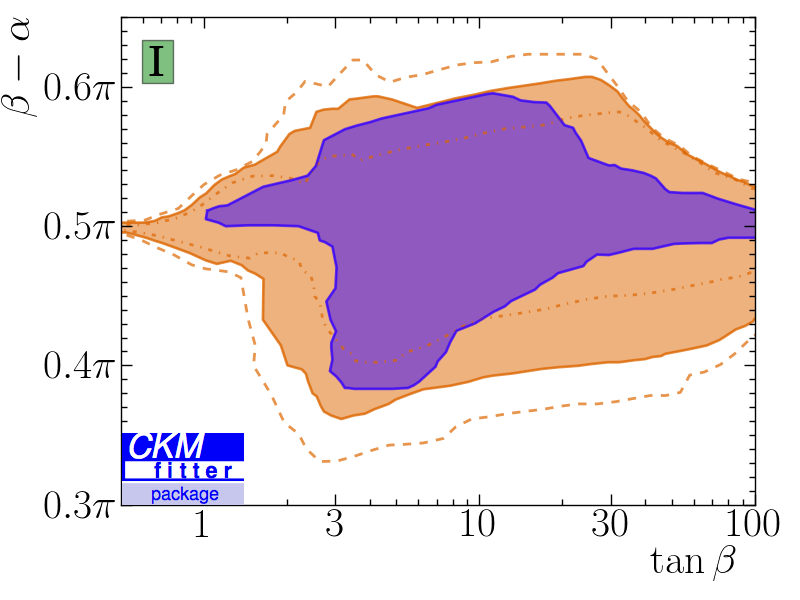}}
    \put(240,165){\includegraphics[width=220pt]{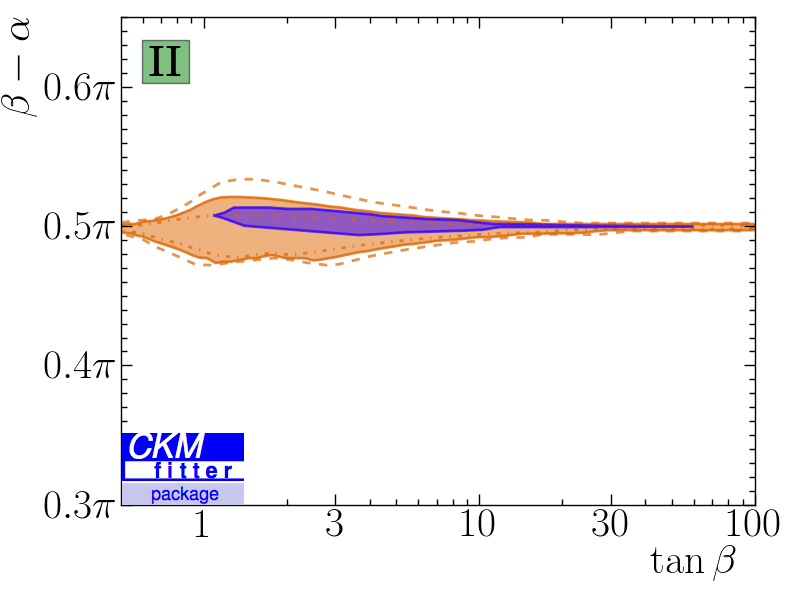}}
    \put(0,0){\includegraphics[width=220pt]{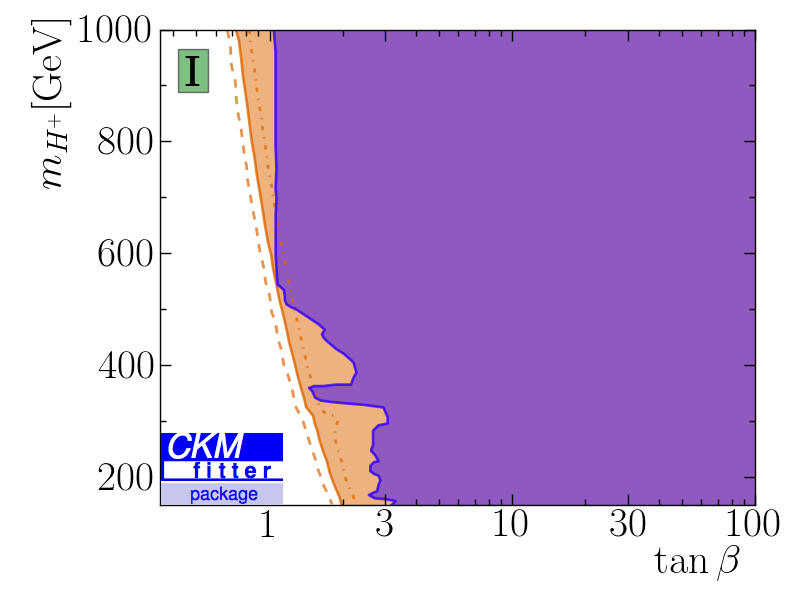}}
    \put(240,0){\includegraphics[width=220pt]{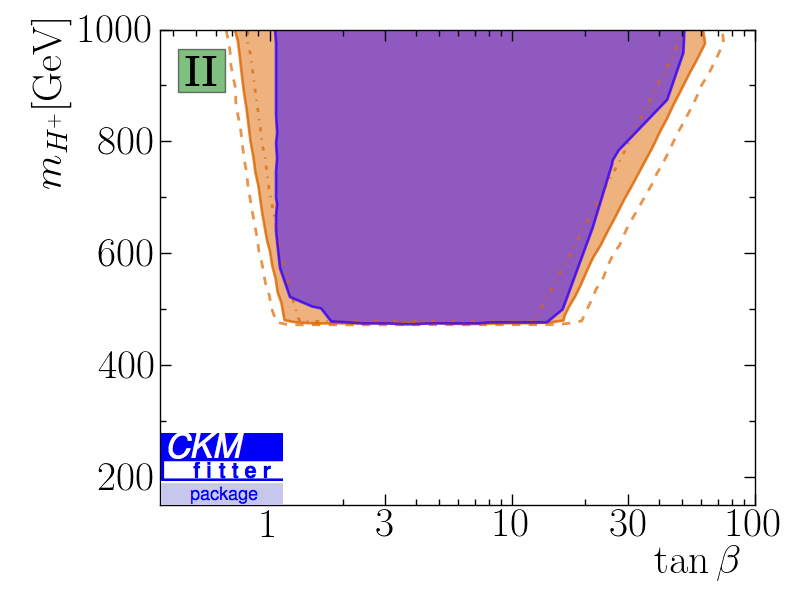}}
   \end{picture}
  }
  \caption{$\tan \beta$--($\beta-\alpha$) and $\tan \beta$--$m_{H^+}$ planes in type I (left) and type II (right) at $m_Z$ with stability imposed at $m_Z$ in orange and at $\mu_{\text{Pl}}$ in purple. The dash-dotted, continuous and dashed lines correspond to the $1\sigma$, $2\sigma$ and $3\sigma$ boundaries, respectively; the $2\sigma$ region -- which roughly corresponds to the $95\%$ C.L. area -- is shaded.}
  \label{fig:logtbvsbmaandmHp}
\end{figure}

The present fit takes into account the latest inputs; therefore we first want to discuss new features before we address RGE related effects.
In Fig.~\ref{fig:logtbvsbmaandmHp}, we show the dependence of $\beta-\alpha$ and $m_{H^+}$ on $\tan \beta$ for type I and II.\footnote{For the other two types we refer to our publication \cite{Chowdhury:2015yja}.}

The $2\sigma$ allowed regions are orange shaded. We can see that between $0.5$ and $100$ all values of $\tan \beta$ are allowed, if the heavy masses ($m_H$, $m_A$, $m_{H^+}$) are not limited to be smaller than $1$ \tev; in the latter case we get a lower bound of $0.7$ from $\Delta m_s$ and -- in type II -- an upper bound of $60$ on $\tan \beta$. This upper bound is a combined effect of various inputs: the EWPO force the neutral heavy masses to be relatively close to $m_{H^+}$, which itself cannot be smaller than $480$ \gev due to the $\oline{B}\to X_s\gamma$ bound. In this mass region however, LHC Higgs data does not allow for too large values of $\tan \beta$. The same effect excludes deviations from the alignment limit $\beta-\alpha=\pi/2$ by more than $0.03\pi$ in type II, which before the $\oline{B}\to X_s\gamma$ update appeared as a second branch in the $\tan \beta$--($\beta-\alpha$) plane.
If we assume additionally that $\mu_{\text{stability}}\geq \mu_{\text{Pl}}$ (purple shaded regions), we obtain a mass-independent lower limit of $\tan \beta>1$. We also observe that $\beta-\alpha$ is constrained even stronger than for $\mu_{\text{stability}}=m_Z$.

\begin{figure}
  \centering
  \resizebox{450pt}{!}{
   \begin{picture}(450,330)(0,0)
    \put(0,165){\includegraphics[width=220pt]{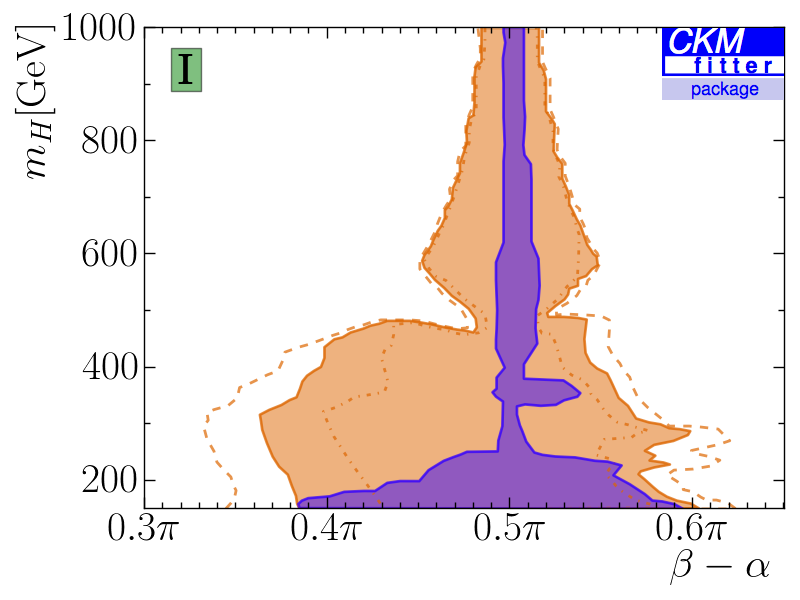}}
    \put(240,165){\includegraphics[width=220pt]{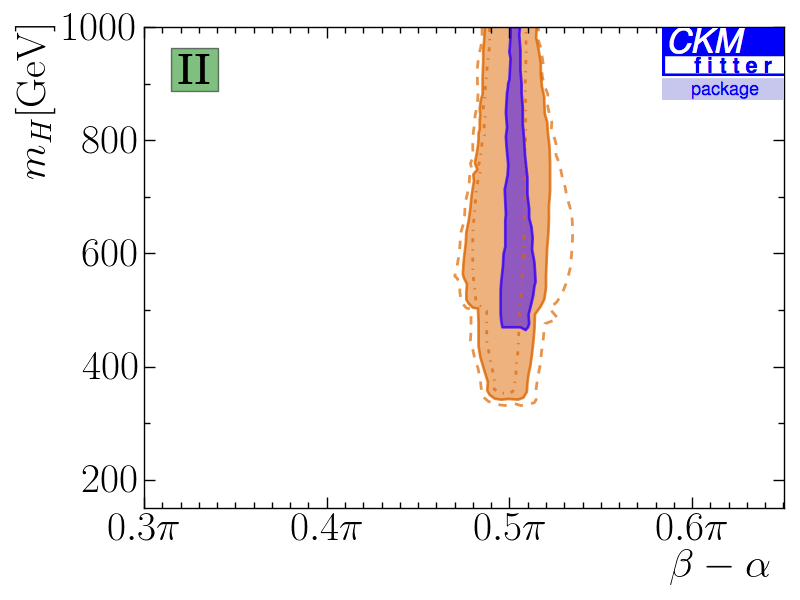}}
    \put(0,0){\includegraphics[width=220pt]{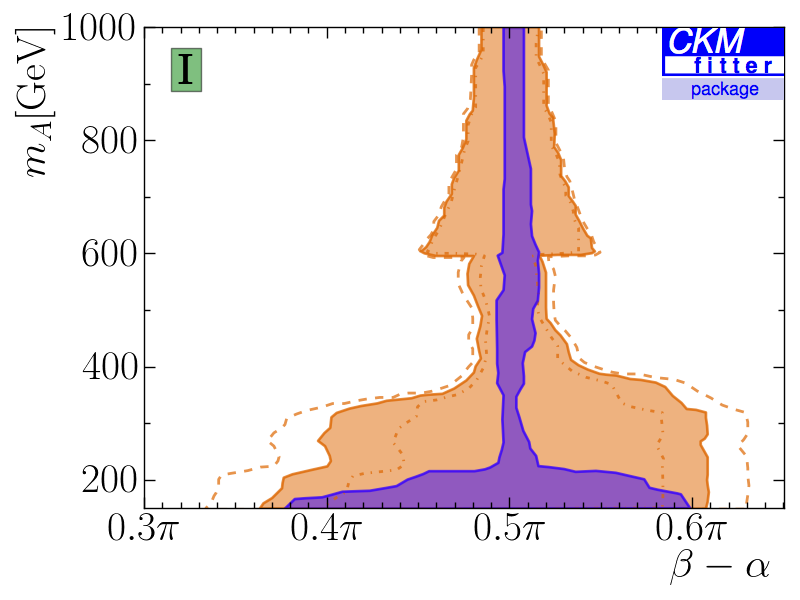}}
    \put(240,0){\includegraphics[width=220pt]{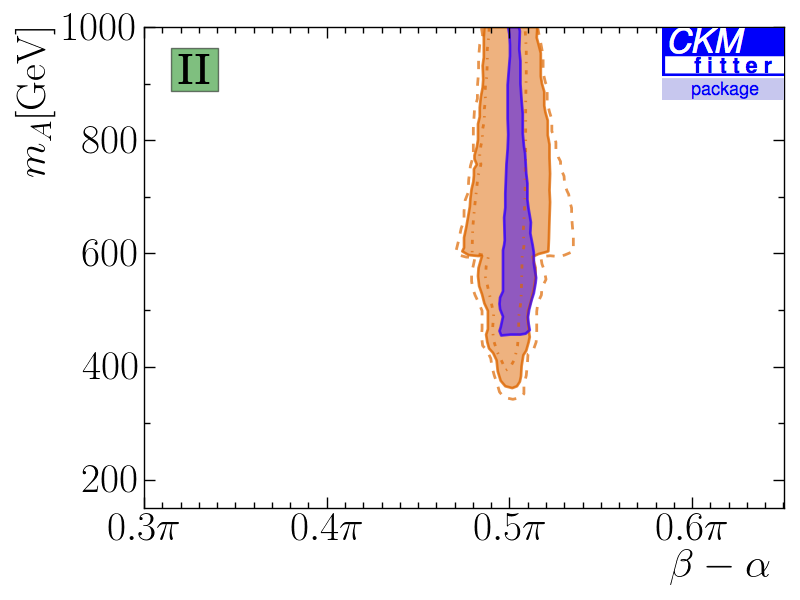}}
   \end{picture}
  }
  \caption{($\beta -\alpha$)--$m_H$ and ($\beta -\alpha$)--$m_A$ planes in type I (left) and type II (right) at $m_Z$ with stability imposed at $m_Z$ in orange and at $\mu_{\text{Pl}}$ in purple. Same colour code as in Fig.~\protect \ref{fig:logtbvsbmaandmHp}.
  }
  \label{fig:bmavsmHandmA}
\end{figure}

In Fig.~\ref{fig:bmavsmHandmA} we quantify the dependence of the maximal $\beta-\alpha$ deviation from the alignment limit on $m_H$ and $m_A$. In type I and for $\mu_{\text{stability}}=m_Z$, sizeable deviations feature $m_H<500$ \gev and $m_A<400$ \gev, where the main constraints on $\beta-\alpha$ stem from the direct heavy Higgs searches.
In type II, these light scenarios are excluded and we obtain lower limits of roughly $350$ \gev for the heavy neutral Higgs masses. These bounds are even increased to $450$ \gev, if we impose stability up to the Planck scale in type II, while in type I larger deviations from the alignment limit are only possible for very light neutral Higgs particles in this case.

\section{The naturalness problem}

After having demonstrated that the 2HDM potential can be stable up to $\mu_{\text{Pl}}$, we want to discuss whether 2HDM scenarios can be found which explain why $m_h$ is in the electroweak regime. (Since we assume $m_H$, $m_A$ and $m_{H^+}$ lower than $10$ \tev, three more natural problems would arise in principle, but here we only want to address the one for the $h$.) If one calculates higher order corrections to quadratic coupling $m_h^2$ in the Lagrangian, the problematic contributions stem from terms quadratic in the cut-off scale $\mu_{\text{nat}}$. The full expression for these corrections $\delta m_h^2$ is

\begin{align*}
 \delta m_h^2 &=
\begin{picture}(50,20)(0,0)\put(-26,-18){\includegraphics[scale=0.75,trim=120 630 10 10,clip=true]{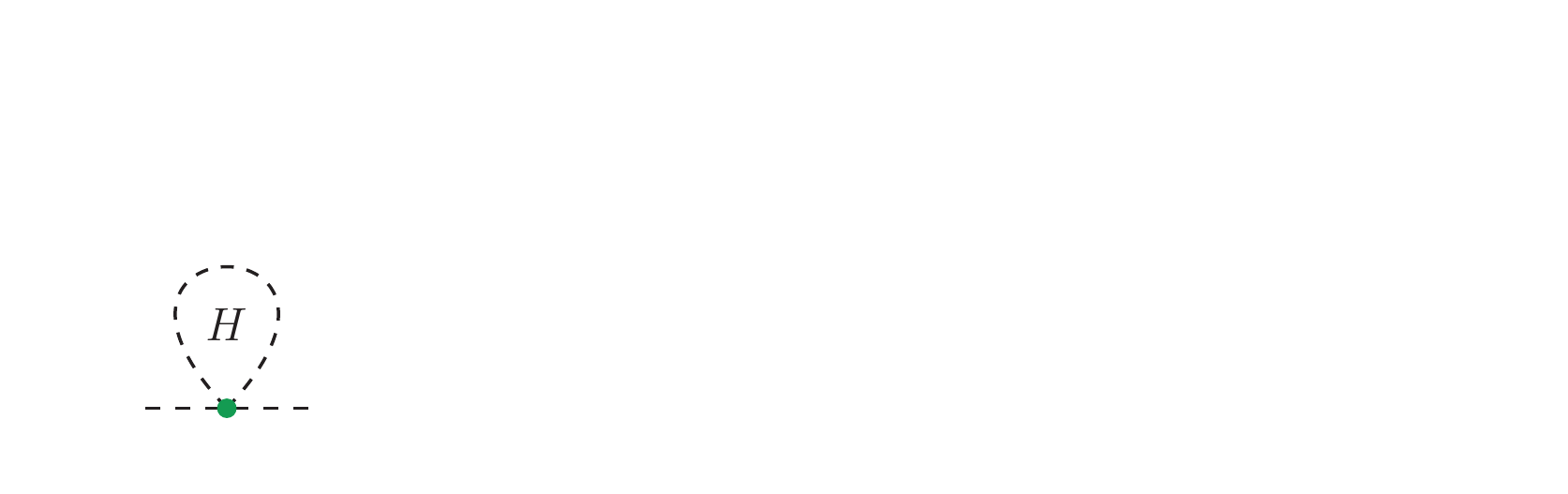}}\end{picture}
+
\begin{picture}(60,20)(0,0)\put(-23.5,-18){\includegraphics[scale=0.75,trim=120 630 10 10,clip=true]{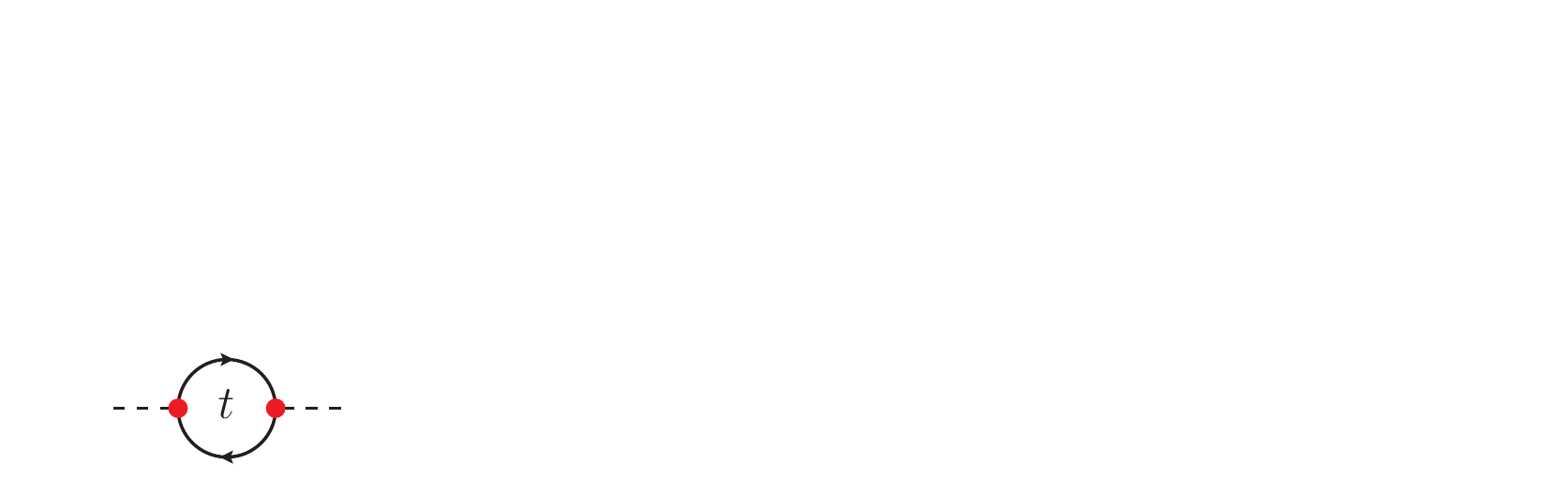}}\end{picture}
+
\begin{picture}(60,20)(0,0)\put(-27,-19){\includegraphics[scale=0.75,trim=145 550 100 10,clip=true]{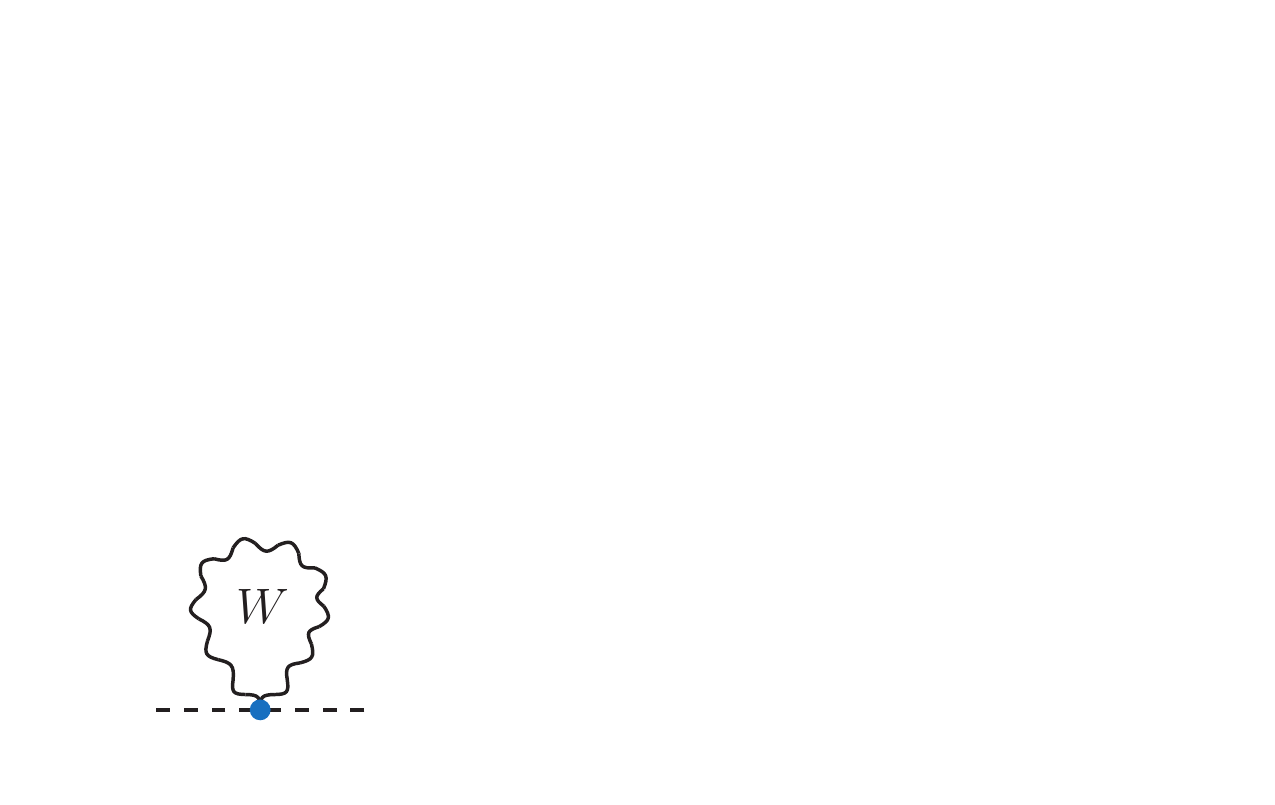}}\end{picture}
+ \; \dots \\[10pt]
 & =\frac{\mu_{\text{nat}}^2}{16\pi^2} \left[ \sum\limits_{n=0}^\infty f_n(\green{\lambda_i},\red{Y_i},\blue{g_i}) \left( \ln \frac{\mu_{\text{nat}}}{\mu_{\text{ew}}}\right) ^n\right] + {\cal O}\left( \ln \frac{\mu_{\text{nat}}}{\mu_{\text{ew}}}\right),
\end{align*}

where the three 2HDM one-loop contributions are shown in the first line and the corrections are split into quadratic and logarithmic terms of $\mu_{\text{nat}}$. The sum is over all loop orders (starting with $n=0$ at one loop), and one can see that in order to cancel all contributions quadratic in the cut-off scale the coefficient functions $f_n(\lambda_i,Y_i,g_i)$ have to vanish. Introducing
 
\begin{align*}
k_n &= \frac{f_{n}(\lambda_i,Y_i,g_i)}{f_{n-1} (\lambda_i,Y_i,g_i)} \ln \frac{\mu_{\text{nat}}}{\mu_{\text{ew}}}
\end{align*}

we can rewrite the quadratic terms

\begin{align*}
 \delta m_h^2 &\approx \frac{\mu_{\text{nat}}^2}{16\pi^2} f_0(\lambda_i,Y_i,g_i) \left[ 1+ \sum\limits_{n=1}^\infty \prod\limits_{\ell=1}^n k_\ell \right].
\end{align*}

This form has the advantage that we can expand the series for small $k_1$ (and, to be sure, also $k_2$), thus cutting after the three-loop contributions. This assumption of perturbativity seems to be arbitrary, however it is the only 2HDM sector for which we can make reliable predictions about the naturalness; in that sense this assumption can be compared to the one of perturbativity of the $\lambda_i$ and $Y_i$ couplings.

In the fit, we take $\mu_{\text{nat}}$ to be the stability cut-off $\mu_{\text{stability}}$, assume that $k_1$ and $k_2$ do not exceed $1$ in magnitude, and impose $|\delta m_h^2 | \leq m_h^2$, which is considered to be natural. Our result is that the absolute value of the one-loop coefficient function $f_0(\lambda_i,Y_i,g_i)$ cannot be larger than $6$ and the maximal perturbative naturalness cut-off is found to be at $5$ \tev. This result holds for all types of $Z_2$ symmetries.

\section{Conclusions}

For all four 2HDM types with a softly broken $Z_2$ symmetry we determine the RGE at NLO with the help of the PyR@TE package; the expressions can be found in the corresponding publication \cite{Chowdhury:2015yja}. Combining these with all relevant up-to-date constraints in a global fit, we find that $\beta-\alpha$ has to be close to the alignment limit in type I and even closer in type II ($\left|\beta-\alpha-\frac{\pi}{2}\right|<0.14\pi$ and $\left|\beta-\alpha-\frac{\pi}{2}\right|<0.025\pi$, respectively). If in addition we demand perturbativity and the stability of the Higgs potential up to the Planck scale, we obtain $\tan \beta>1$ in both types and even stronger bounds on $\beta-\alpha$. Furthermore, we analyse the hierachy problem and -- assuming a suppression of higher order $m_h$ corrections -- find that the light Higgs mass can only be explained naturally if the cut-off scale is in the \tev range.

\bibliographystyle{utphys}
\bibliography{lambdathdm_proc}

\end{document}